\documentclass[preprint]{aastex}
\usepackage{graphics}
\usepackage{epsfig}
\usepackage{subfigure}

\newcommand{\smyr}{M$_\odot$ yr$^{-1}$}
\newcommand{\vkm}{km s$^{-1}$}
\newcommand{\degree}{$^{\circ}$}

\def\includegraphics#1{#1}

\def\leftblank#1{}
\def\degree{$^{\circ}$}

\def\vkm{km s$^{-1}$}
\def\smyr{M$_\odot$ yr$^{-1}$}

\def\cm3{cm$^{-3}$}
\def\arcs#1{$#1''$}
\def\arcsa#1#2{$#1^{\prime\prime}_{^\textrm{.}}#2$}

\def\H2{H$_2$}

\def\putfig#1#2#3{\epsfig{scale=#1,angle=#2,figure=#3}}

\def\leftblank#1{}

\begin{document}
\title{Collimated Fast Wind in the Pre-Planetary Nebula CRL 618}
\author{Chin-Fei Lee\altaffilmark{1}, Ming-Chien Hsu\altaffilmark{1},
and Raghvendra Sahai\altaffilmark{2}}
\altaffiltext{1}{Academia Sinica Institute of Astronomy and Astrophysics,
P.O. Box 23-141, Taipei 106, Taiwan; cflee@asiaa.sinica.edu.tw}
\altaffiltext{2}{
Jet Propulsion Laboratory, MS 183-900, 4800 Oak Grove Drive,
Pasadena, CA 91109
}

\begin{abstract}
Collimated fast winds (CFWs) have been proposed to operate during the 
post-AGB evolutionary phase (and even earlier during the late AGB phase),
responsible for the shaping of pre-planetary nebulae (PPNs)
and young planetary nebulae (PNs).
This paper is a follow-up to our previous study of CFW
models for the well-studied PPN CRL 618. 
Previously, we compared our CFW models
with optical observations of CRL 618 in atomic and ionic lines and found that
a CFW with a small opening angle can readily reproduce the highly collimated
shape of the northwestern (W1) lobe of CRL 618 and the bow-like structure seen at its tip.
In this paper, we compare our CFW models with recent
observations of CRL 618 in CO J=2-1, J=6-5, and \H2{} 1-0 S(1).
In our models, limb-brightened shell structures are seen in CO and \H2{} at
low velocity arising from the shocked AGB wind in the shell, and can be
identified as the low-velocity (LV) components in the observations.
However, the shell structure in CO J=2-1 is significantly less
extended than that seen in the observations.  None of our models can
properly reproduce the observed high-velocity (HV) molecular emission near the source
along the body of the lobe. In order to reproduce the HV molecular
emission in CRL 618, the CFW is required to have a different structure. One
possible CFW structure is the cylindrical jet, with the fast wind material
confined to a small cross section and collimated to the same direction along
the outflow axis. \end{abstract}


\section{Introduction}

At the end of the evolution of low- and intermediate-mass stars,
pre-planetary nebulae (PPNs) are seen associated with post asymptotic giant
branch (post-AGB) stars. They will turn into planetary nebulae (PNs) in less
than 1000 years as the post-AGB stars become hot white dwarfs.  Their
shaping mechanism is still unclear and is closely related to the mass-loss
processes during the end phases of the evolution.

Many PPNs and young PNs were found to have highly aspherical shapes, with a
significant fraction having highly collimated bipolar or multipolar lobes
\citep{Corradi1995,Schwarz1997,ST98,Sahai01,Sahai2007}.
Point symmetry, rather than
axisymmetry, better characterizes the geometry of the majority of these
objects [striking examples are He\,2-115 \citep{ST98}
and He\,3-1475 \citep{Borkowski97}].
As a result,
instead of spherical fast winds as in the
generalized interacting stellar winds (GISW) model 
\cite[see e.g., review by][]{Balick2002},
collimated fast winds have been proposed to operate during the 
post-AGB phase (and even earlier during the late AGB phase),
and be the primary agents for the shaping of PPNs and young PNs \citep{ST98,Sahai01}.
Collimated fast wind (CFW) models have been used to account for the
morphology and kinematics of a few well-studied PPNs and 
PNs, with some assuming a radial wind with a small opening angle
\cite[Lee \& Sahai 2003, hereafter \citet{LS03};][]{Akashi2008}, some 
assuming a cylindrical jet either unmagnetized \citep{Cliffe95,Steffen98,Guerrero2008}
or magnetized \citep{LS04}, and some assuming a bullet (a massive clump)
along the outflow axis \citep{Dennis08}.

This paper is a follow-up to our previous study of CFW
models for the well-studied PPN CRL 618 \citep{LS03}.
CRL 618 is located at a distance of 900 pc
\citep{Goodrich1991}. It
shows several narrow lobes at different orientations in HST images 
\citep{TG02} and belongs to the ``multipolar"
morphological classification \citep{Sahai2007},
perhaps resulting from multiple ejections at different orientations.
The general structures of the different lobes 
are similar and we focus only on the northwestern (W1) lobe, 
which seems to be better separated from other lobes.
In \citet{LS03}, we compared our CFW models
with optical observations of CRL 618 in atomic and ionic lines and found that
a CFW with a small opening angle can readily reproduce the highly collimated
shape of the W1 lobe
and the bow-like structure seen at its tip
\citep{Sanchez2002}. However, it may have difficulties in reproducing
properly the
high-velocity optical emission along the body of the lobe.
In this paper, we compare our CFW models with recent observations of CRL 618 in \H2{}
1-0 S(1) \citep{Cox03}, CO J=2-1 \citep{Sanchez2004} and J=6-5
\citep{Nakashima07}. We find that our CFW models also have difficulties in
reproducing the high-velocity molecular emission in the W1 lobe and the
CFW is required to have a different physical structure.

\section{Numerical Methods and Physical Settings}

The two-dimensional hydrodynamic code, ZEUS 2D, is used
for the simulations of our models, as in \citet{LS03}.
This code has been now modified to include molecular cooling and
the time dependent chemistry of 
hydrogen by solving the following equations \cite[see also][]{Suttner1997}:
\begin{eqnarray}
\frac{\partial e}{\partial t} + \nabla(e \cdot \mathbf{v}) 
&=& -p\nabla \cdot \mathbf{v} - \Lambda_A(T,n,f,g)-\Lambda_M(T,n,f,g)  \\
\frac{\partial(fn)}{\partial t} + \nabla(fn \cdot \mathbf{v})
&=& R(T,n,f,g) - D(T,n,f,g) \\
\frac{\partial(gn)}{\partial t} + \nabla(gn \cdot \mathbf{v})
&=& I(T,n,f,g) - C(T,n,f,g)
\end{eqnarray}
where $e$, $p$, $\mathbf{v}$, and $T$ are the internal energy density,
thermal pressure, velocity, and temperature, respectively.
Also, $n$ is the hydrogen nuclei density, $f$ is the fractional part of
hydrogen molecules with $n_{\scriptsize\textrm{H$_2$}}=f n$ (i.e.,
$f=0$ for atomic/ionic gas and $f=0.5$ for molecular gas), 
and $g$ is the fractional part of ionized hydrogen with $n_{\scriptsize
\textrm{H$^+$}}=g n$. Helium is included as a neutral component with
$n_{\scriptsize \textrm{He}}=0.1 n$ and thus $n=\rho/(1.4m_{\scriptsize
\textrm{H}})$, where $\rho$ and 
$m_{\scriptsize \textrm{H}}$ are the mass density and the mass of atomic
hydrogen, respectively.
Here $D$ and $R$ are dissociation and ressociation rates of molecular hydrogen,
respectively \cite[see][and reference therein]{Suttner1997}, $I$ and $C$ are ionization
and recombination rates of atomic hydrogen, respectively.
$\Lambda_{A}$ is the optically thin
radiative cooling from atoms and ions \cite[as in][]{LS03}, with
the cooling at high temperature from \citet{MacDonald1981} and
low temperature from \citet{Dalgarno1972}.
$\Lambda_{M}$ is the optically thin radiative cooling from
molecules, including \H2{} \citep{Hollenbach1979} 
and CO \citep{Mckee1982,Hollenbach1989}.
The chemistry of CO is not included.
The CO abundance is assumed to be constant and equal to
2$\times10^{-4}$ of the number of hydrogen molecules, as in \citet{Sanchez2004}.
We believe this assumption does not significantly affect our conclusions, 
it only affects the small-scale structure of the CO emission.
The equations of 
state and material coefficients are \cite[for similar derivation see][]{Suttner1997}:
\begin{eqnarray}
p&=&\frac{\rho k T}{\mu m_{\scriptsize \textrm{H}}} \hspace{2cm}
e=\frac{\rho C_v k T}{m_{\scriptsize \textrm{H}}} \nonumber \\
\mu&=&\frac{1.4}{1.1-f+g} \hspace{0.8cm} C_v =\frac{3.3-f+3g}{2.8}
\end{eqnarray}
where $k$, $\mu$, and $C_v$ are the Boltzmann constant, mean molecular
weight, and specific heat, respectively.
A scalar color tracer $c$ is also included in the simulations to track
the fast wind,
it is one for the fast wind,
zero for the AGB wind, and a value between one and zero 
for a mixture of the fast wind and AGB wind.
As in \citet{LS03}, the simulations are performed in spherical coordinates
but presented in cylindrical coordinates (z, R), with the z-axis being the outflow axis.

Our CFW models are based on models 1 and 4 in \citet{LS03}, which were found
to be the best models for the W1 lobe of CRL 618 as seen at optical
wavelengths. In these models, a CFW with a small opening angle is assumed to
emanate radially from the vicinity of the post-AGB star, interacting with a
spherical AGB wind \cite[for details see][]{LS03}. In our simulations,
the CFW is launched
from the inner boundary of the simulation domain, which is
at 5$\times10^{15}$ cm ($\sim$ 333 AU) away from the post-AGB star.
The AGB wind is assumed to be molecular with a temperature of 10 K and have a
mass-loss rate of $3\times10^{-5}$ \smyr{} with an expansion velocity of
$20$ \vkm{}. The CFW is assumed to have a mass-loss rate of
2.5$\times10^{-7}$ \smyr{}, a speed of 300
\vkm{}, and an opening angle of $10^\circ$.
It can either be steady or pulsed with a temporal variation in the
density and velocity (see Table \ref{tab:models}).
It can be either atomic or molecular, depending on the temperature of the
CFW to be assumed.

\section{Recent Observations in CO and \H2{}}

CRL 618 has been recently observed in \H2{} 1-0 S(1) at $\sim$ \arcsa{0}{5}
resolution \citep{Cox03} and in CO J=2-1 \citep{Sanchez2004} and J=6-5
\citep{Nakashima07} at $\sim$ \arcs{1} resolution (see Figs.
\ref{fig:momobs} and \ref{fig:pvobs}). It is multipolar, but here we focus
only on its W1 lobe that
extends $\sim$ \arcs{6} to the west from the source.
The systemic velocity in this region is $-$21.5 \vkm{} LSR and the
W1 lobe is mainly redshifted with a velocity ranging from -40 to 150 \vkm{} LSR. 
The CO and \H2{}
emission toward the W1 lobe can be roughly separated into two components: a
slow or low-velocity (LV) component with a velocity lower than $\sim$ 20$-$30 \vkm{}
from the systemic and a fast or high-velocity (HV) component with a velocity 
extending to $\sim$ 170 \vkm{} from the systemic. The
LV component is extended. In CO J=2-1, it forms a limb-brightened shell
structure around the optical lobe, extending to $\sim$ \arcs{5} away from
the source. In CO J=6-5, it also forms a shell structure but only in the
southern part of the lobe, extending to only $\sim$ \arcs{3} away from the
source. As discussed in \citet{Nakashima07}, however, it should be more
extended, because most of the extended flux has been resolved out in their
observations. In \H2{}, it is spatially unresolved, extending mainly from
\arcs{2} to \arcs{6} from the source, slightly ahead of that seen in CO
J=2-1. On the other hand, the HV component is more compact. In CO J=2-1, it extends to $\sim$
\arcsa{2}{5} away from the source with the velocity increasing linearly with
the distance from near the systemic velocity to the highest velocity. In
\H2{}, it is seen with three localized emission peaks separated by $\sim$
\arcs{2}: (1) a peak close to the source with a range of 
(blueshifted and redshifted) velocities,
(2) a bright peak at $\sim$ \arcsa{2}{5}, where the
tip of the CO HV component is, with a broad range of velocities extending
from near the systemic velocity to the highest velocity, and (3)
a faint peak at $\sim$ \arcs{4} at $\sim$ 80 \vkm{} LSR.

\section{Results}

In the following, we present our models and
the comparison with the observations.
Figure \ref{fig:models} shows the distributions of hydrogen nuclei density
and temperature with molecular fraction 
in our models at the age of $\sim$ 160 yrs, when the outflow
lobe has a length of $\sim$ 6000 AU, similar to the deprojected length of
the W1 lobe of CRL 618.
The separation between the AGB wind
and the CFW is delineated by the color tracer $c=0.5$, at which half is the
AGB wind and half is the CFW.
In order to compare with the observations, we also
derive intensity maps of the LV and HV components (Fig.
\ref{fig:modelmoms2com}) and position velocity (PV) diagrams (Fig.
\ref{fig:modelpvs}) for the CO J=2-1, J=6-5, and \H2{} 1-0 S(1) emission
from our models, assuming the latter to be optically thin and arising
from gas in local thermal equilibrium (LTE). We assume a distance of 900 pc
and an inclination of 30\degree{}, values appropriate for CRL 618
\citep{Sanchez2004}. The HV components are derived by integrating the
emission with velocity higher than 50 \vkm{} from the systemic and the LV
components by integrating the emission with velocity within 50 \vkm{} from
the systemic.  Two angular resolutions, \arcsa{0}{1} and \arcsa{0}{5}, are
assumed, with the latter for comparing with the current observations.

\subsection{Model 1: Steady Atomic CFW at 10,000 K}

\begin{figure}
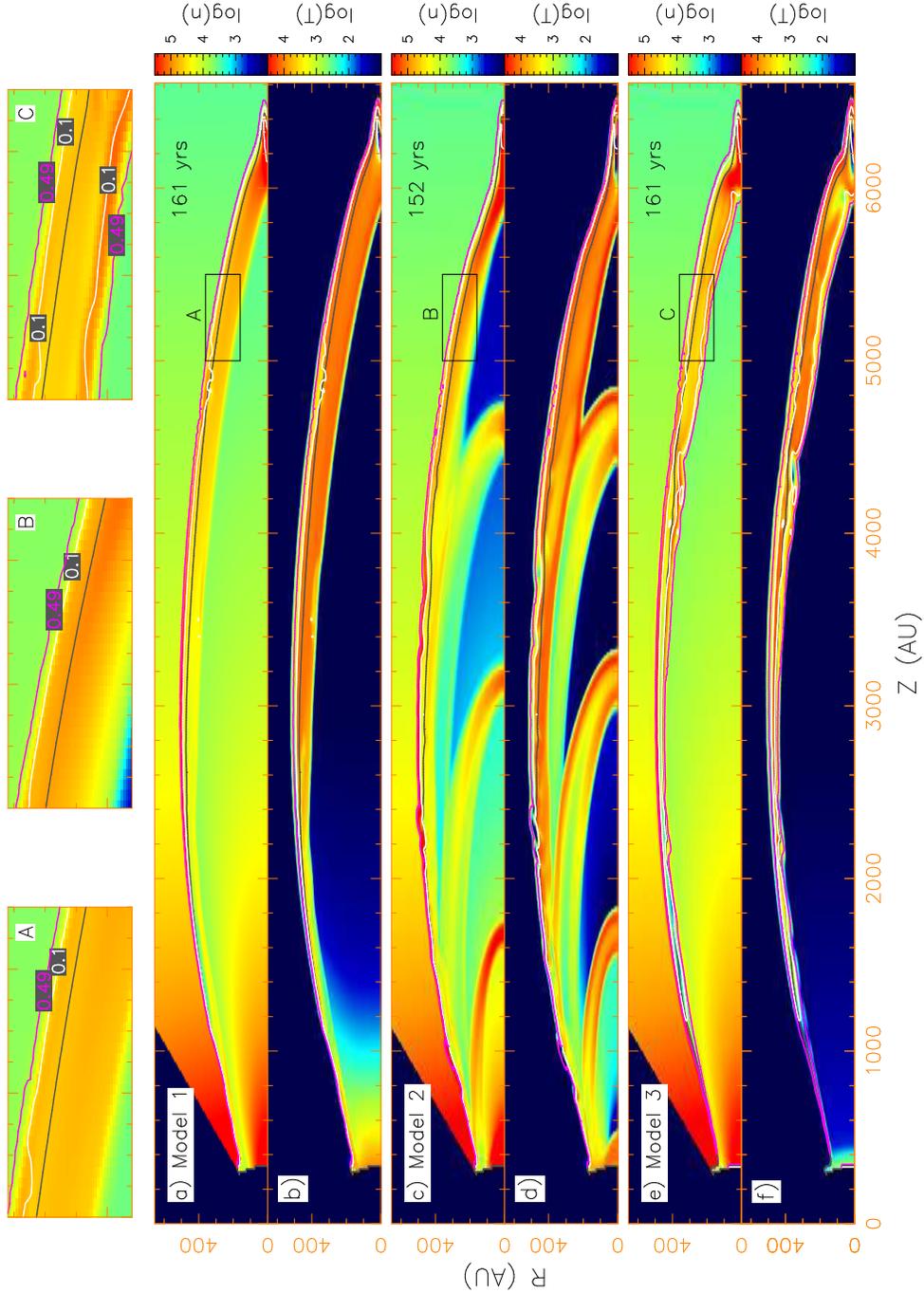

\centering 
\putfig{0.75}{0}{f3a.ps}
\figcaption[] 
{Simulations of our CFW models from 1 to 6, at the age of $\sim$ 160 yrs.
(a), (c), (e), (g), (i), and (k) show the number density 
of hydrogen nuclei in logarithmic scale.
(b), (d), (f), (h), (j), and (l) show the temperature in logarithmic scale.
The molecular fractions $f=0.1$ (white line) and $f=0.49$ (magenta line) 
are shown to outline the distribution of molecular gas.
The separation between the AGB wind and the CFW is shown by the
color tracer $c=0.5$ (gray line, at which half is AGB wind and
half is CFW).
The CFW is atomic at 10,000 K in models 1 and 2, while is molecular at 1000 K
in models 3, 4, 5, and 6. The CFW is steady in models 1 and 3, while is
pulsed with A=0.5 in models 2 and 4, A=0.3 in model 5, 
and A=0.1 in model 6.
(A)$-$(F) show respectively the blow-ups for the regions in the boxes
A$-$F.
\label{fig:models} }
\end{figure}

Model 1 here corresponds to model 1 in \citet{LS03}. The CFW has a
temperature of 10$^4$ K and is thus atomic. As it blows into the AGB wind,
it produces a collimated outflow lobe, which is a thin shell with a cavity
(Fig. \ref{fig:models}a). As shown, the shell consists of shocked AGB wind
in the outer shell and shocked fast wind in the inner shell (Fig.
\ref{fig:models}A). Since the shell
is already radiative and thus momentum-driven even without molecular
cooling, additional molecular cooling does not change the shell dynamics, it
only reduces the shell thickness as compared to that in \citet{LS03}. In the
shell, the shock becomes stronger going from near the source toward the tip
of the lobe.  As a result, the temperature of the shock AGB wind increases
from tens K near the source to above 10$^4$ K near the tip (Fig.
\ref{fig:models}b).  In addition, the shocked AGB wind is mainly molecular
near the source but becomes mainly atomic near the tip (except for the newly
shocked AGB wind) because of shock dissociation. In contrast, the shocked
fast wind is mainly atomic.

In the simulation, a jet-like structure is seen at the tip because of the
accumulation of material there due to the shock focusing effect, as
discussed in \citet[][]{LS03}. Such accumulation might be artificially
enhanced in our 2D simulations because material cannot flow across the
symmetry axis. This enhancement in turn would lead to more cooling and
further accumulation, and then an increase in the formation rate of
molecules. Therefore, the molecular emission from the lobe tip might be
highly overestimated and should be kept in mind when comparing our models to
the observations.

\begin{figure}[!hbp]
\centering
\putfig{0.63}{0}{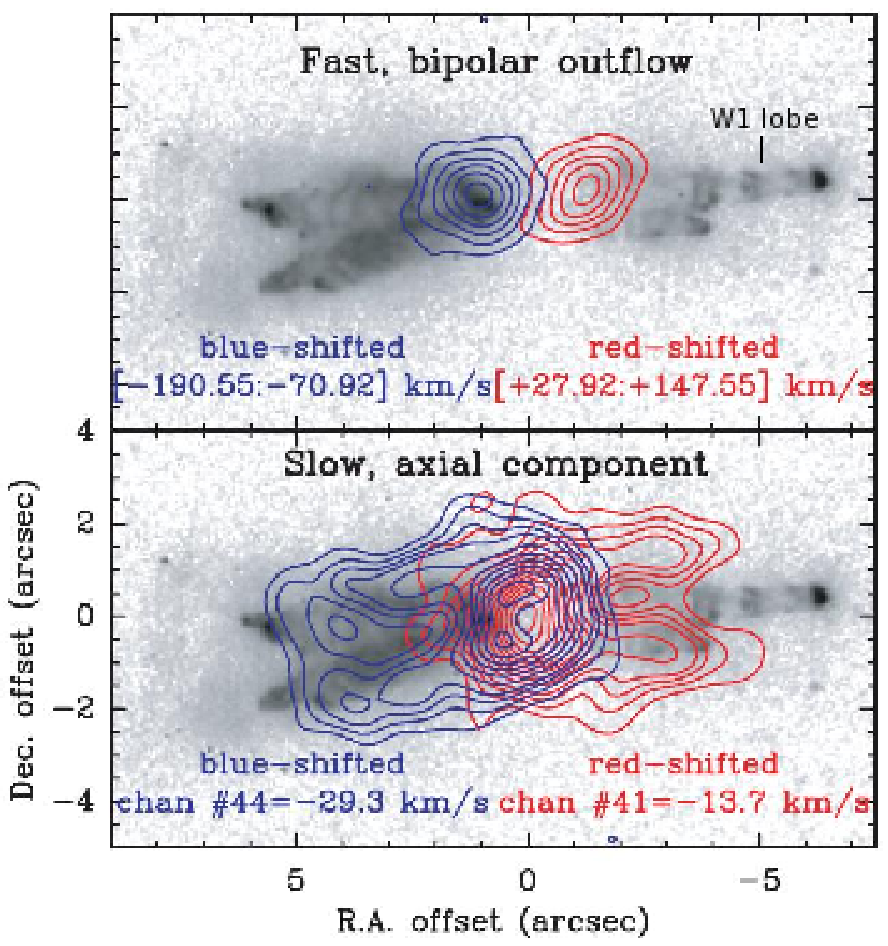}
\hspace{0.01in}\vspace{-0.05in}
\putfig{0.31}{0}{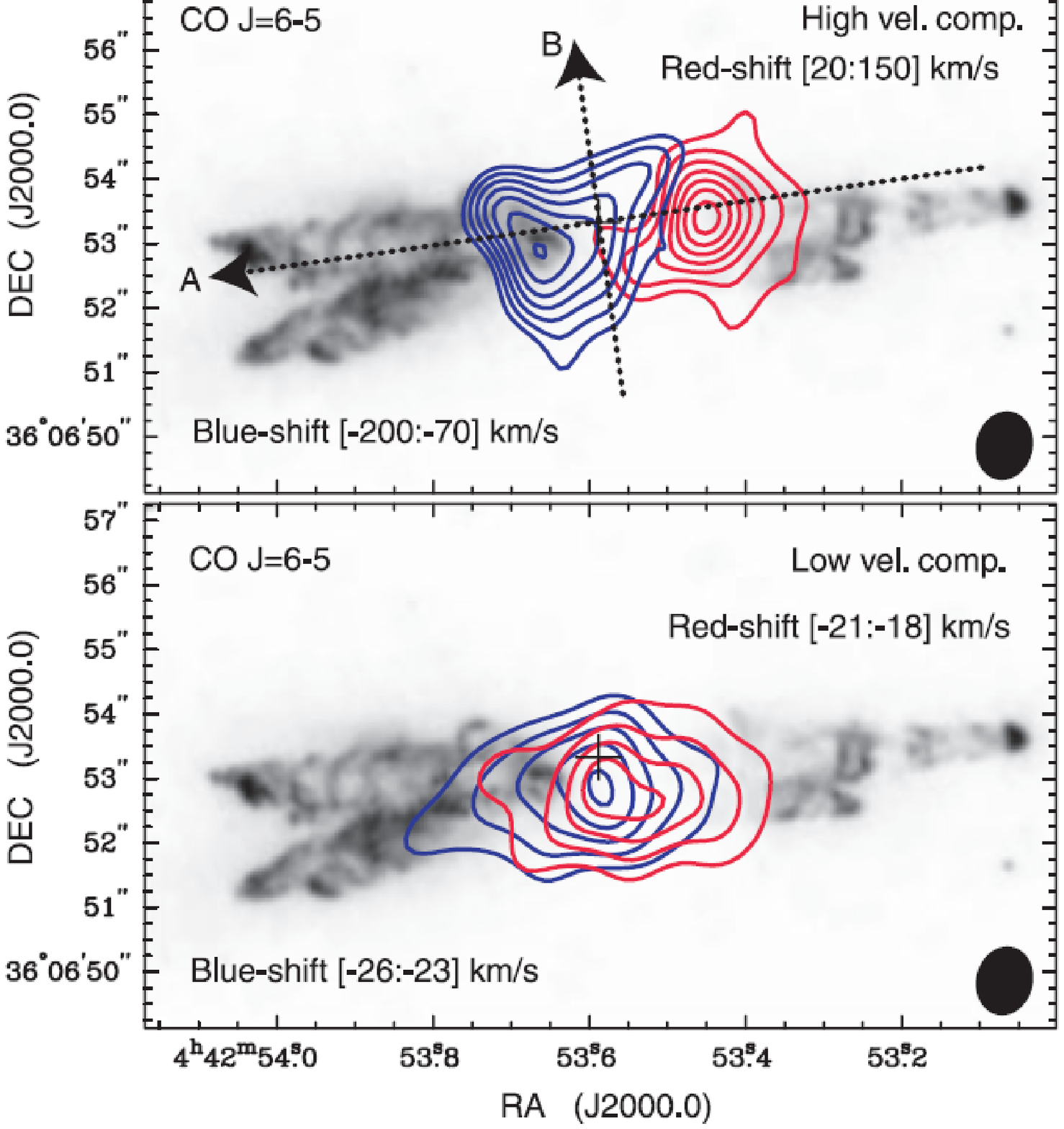}
\putfig{0.63}{0}{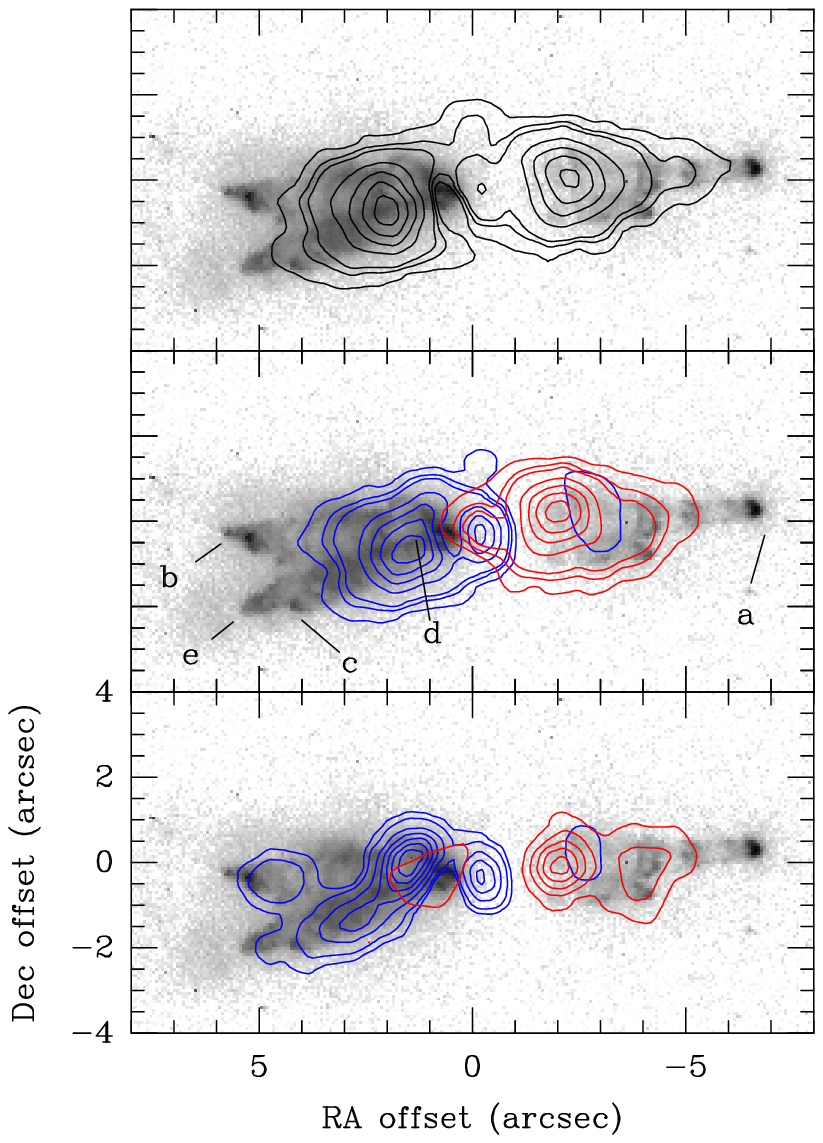}
\figcaption[]
{CO J=2-1, CO J=6-5, and \H2{} 1-0 S(1) emission contours
superposed on the HST optical image of CRL 618.
(Left Column) The top and bottom panels show the contours of
CO J=2-1 emission at high and low velocities, respectively \citep{Sanchez2004}.
(Middle Column) The top and bottom panels show the contours of
CO J=6-5 emission at high and low velocities, respectively \citep{Nakashima07}.
(Right Column) The top, middle, and bottom panels show the contours of
\H2{} 1-0 S(1) emission at low velocity,
middle velocities, and high velocities, respectively \citep{Cox03}. The
source is either at the position offset (0,0) or marked by a cross. 
The W1 lobe is the optical lobe extending
$\sim$ \arcs{6} to the west from the source, as indicated.
\label{fig:momobs}
}
\end{figure}

In this model, the LV components of the CO J=2-1, J=6-5, and \H2{} emission
are seen forming limb-brightened shell structures around the cavity (Fig.
\ref{fig:modelmoms2com}, model 1), as in the CO observations (Fig.
\ref{fig:momobs}). They arise from the shocked AGB wind in the shell. Since
the temperature in the shell increases with the distance from the source,
the shell structure extends further and further away from CO J=2-1 to CO
J=6-5 and to \H2{}. The CO J=2-1 emission traces the cold (20-100 K) gas
extending to $\sim$ \arcs{2} from the source, the CO J=6-5 emission traces
the warm gas (50-300 K) extending to $\sim$ \arcs{3}, and the \H2{} emission
traces the hot gas ($\sim$ 2000 K) extending from \arcs{3} to \arcs{5}.  In
CO J=2-1, however, the shell structure in this model is significantly less
extended than that seen in the observations (Fig. \ref{fig:momobs}),
indicating that the temperature in the shell in this model must have
decreased less rapidly from the tip to the source than that in the
observations.  No HV component is seen in this model due to the lack of
molecules at high velocity, except for the jet-like structure at the tip of
the lobe.

\begin{figure}
\centering 
\putfig{0.54}{0}{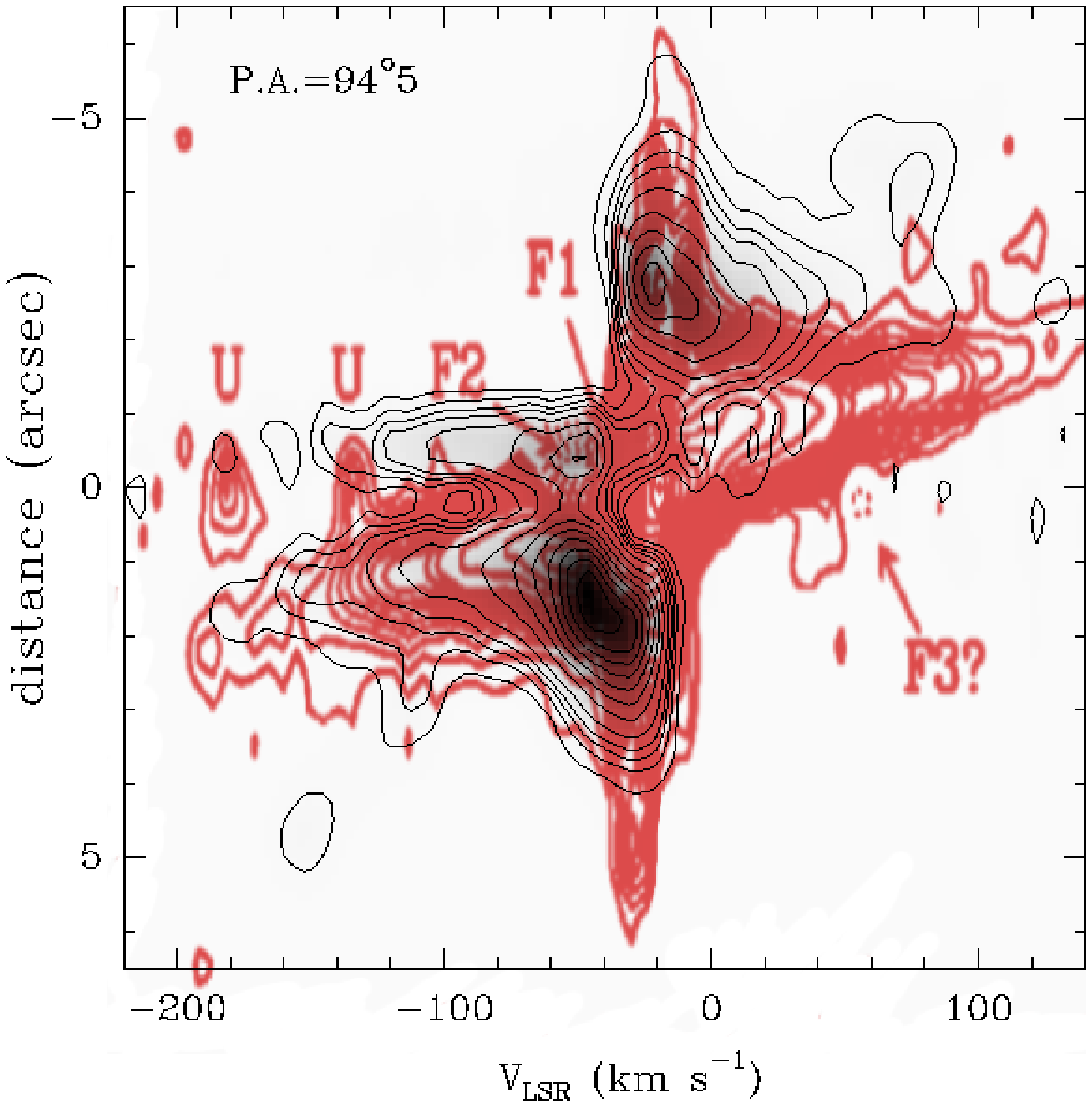}
\putfig{0.54}{0}{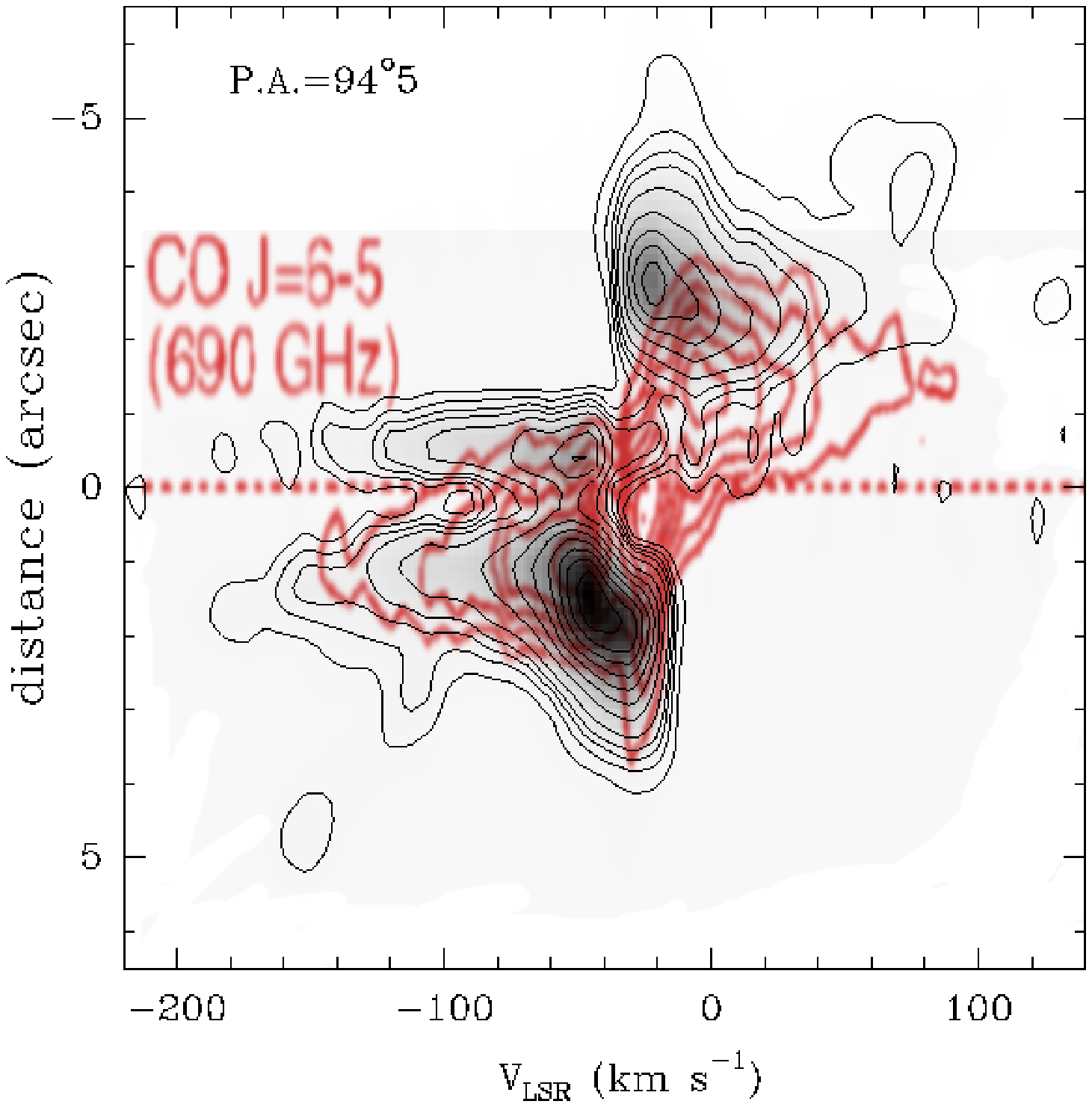}
\figcaption[]
{Position-velocity (PV) diagrams of \H2{} 
\cite[black contours with gray image in both panels,][]{Cox03},
CO J=2-1 \cite[red contours in the left panel,][]{Sanchez2004},
and CO J=6-5 \cite[red contours in the right panel,][]{Nakashima07}
emission cut along the outflow axis.
The W1 lobe is mainly redshifted with a velocity ranging from
$-$40 to 150 \vkm{}
LSR.
\label{fig:pvobs} }
\end{figure}

PV diagrams of CO and \H2{} emission cut along the
outflow axis together show a V-shaped PV structure extending to $\sim$ 40
\vkm{} from the systemic, with the CO emission at the lower end and the
H$_2$ emission at the upper end (Fig. \ref{fig:modelpvs}, model 1).  This PV
structure is associated with the LV components, arising from the shocked AGB
wind in the shell, which is in the expanding lobe. This PV structure is
expected,
with the left and right parts from the back
and front walls of the expanding lobe, respectively. In observations, the LV component
in CO J=2-1 also shows a hint of a V-shaped PV structure (Fig.
\ref{fig:pvobs}) and has also been modeled with an expanding lobe by
\citet{Sanchez2004}. The expansion velocity in their model, however, is
$\sim$ 22 \vkm{}, much lower than that in our model.
The upper end of the V-shaped PV structure is also expected to be seen
in \H2{} even at low resolutions in the observations (Fig.
\ref{fig:modelpvs}, model 1). However, no such PV structure is seen in 
\H2{} observations (Fig. \ref{fig:pvobs}). 
It is likely because the outflow lobe in CRL 618 indeed has a 
smaller expansion velocity as found in the CO J=2-1 observations,
so that the V-shaped PV structure 
could have been smeared out in the observations.

\subsection{Model 2: Pulsed Atomic CFW at 10,000 K}

Model 2 here corresponds to model 4 in \citet{LS03}. The CFW here is the
same as that in model 1 but with a temporal variation in density and
velocity with an amplitude $A=0.5$ (i.e., $\triangle v = 150$ \vkm{}) and a
period $P=22$ yr \cite[see][]{LS03}. In this model, a series of
internal shock pairs are
formed in the cavity moving along the outflow axis at high velocity, as the
faster CFW catches up with the slower CFW (Figs. \ref{fig:models}c, d). They
are mainly atomic. They do not affect the shell dynamics. They only produce
small disturbances (ripples) on the shell structure in the regions where the
internal shocks interact with the shell. Thus, the emission structures in CO
and \H2{} are the same as in model 1, except that there are faint ring-like
structures seen across the cavity arising from the ripples in the shell
(Fig. \ref{fig:modelmoms2com}).  The PV diagrams are also similar to those
in model 1, except that there are faint low-velocity \H2{} emission near the
source arising from the ripples in the shell (Fig. \ref{fig:modelpvs}). No
HV component is seen from the internal shocks due to the lack of molecules.
As a result, although this model was found to produce optical emission in
the cavity with the internal shocks \citep{LS03}, it can not produce
molecular emission in the cavity.

\subsection{Model 3: Steady Molecular CFW at 1000 K}

This model is the same as model 1 but with the CFW assumed to be molecular
at 1000 K, in order to have HV molecular gas inside the cavity.  Note that
the assumed value of the temperature is not important,
because the temperature of the CFW will drop rapidly below 100 K in 200 AU 
(or $\sim$ \arcsa{0}{2})
due to radial expansion and radiative cooling (Fig. \ref{fig:models}f).
Since the shell is not pressure-driven, reducing the temperature of the CFW
does not change the shell dynamics (Fig. \ref{fig:models}e).
Thus, the molecular fraction of the shocked AGB wind in the shell is the same as in
model 1, and so are the emission and PV structures of the LV components
(Figs. \ref{fig:modelmoms2com} and \ref{fig:modelpvs}).
On the other hand, only the newly shocked fast wind in the inner boundary of
the shell is molecular (as indicated by the contours of $f$, Fig.
\ref{fig:models}C)
since the shock there is weak. It is hot,
producing two HV shell structures in \H2{} at the far end, one at
$\sim$ \arcsa{3}{5} around $\sim$ 60 \vkm{} and one at $\sim$
\arcs{5} around $\sim$ 100
\vkm{}, with the velocity increasing with the distance from the source. The
one at $\sim$ \arcsa{3}{5} may have a counterpart at $\sim$ \arcs{4}
in the \H2{} observations (see
Figs. \ref{fig:momobs} and \ref{fig:pvobs}). Since the temperature of the
CFW itself drops below 100 K in 200 AU, HV CO emission is seen arising from
the CFW itself near the source. However, due to radial expansion of the CFW
itself, this HV CO emission is very faint as compared with the LV component
in the PV diagrams
at the resolutions of the observations (Fig. \ref{fig:modelpvs}),
inconsistent with the observations, 
in which the HV CO component is comparable to the LV
component (Fig.  \ref{fig:pvobs}).
Besides, no HV \H2{} emission is seen from the CFW itself near
the source.

\subsection{Model 4: Pulsed molecular CFW at 1000 K}

This model is the same as model 3 but with the CFW assumed to have
a temporal variation in density and
velocity with an amplitude $A=0.5$ (i.e., $\triangle v = 150$ \vkm{})
and a period $P=22$ yr.
As in model 2, the shell structure and dynamics
are not affected much by the internal shocks (Figs. \ref{fig:models}g, h).
Thus, the emission and PV structures of the LV components,
which arises from the shocked AGB wind in the shell, 
are also similar to those in model 2.
The internal shocks, as they propagate beyond 1000 AU ($\sim$ \arcs{1}) from
the source, become strong enough to
dissociate the molecules (Fig. \ref{fig:models}D).
Therefore,
only the internal shock closer to $\sim$ \arcs{1} produces HV \H2{}
emission (Figs. \ref{fig:modelmoms2com} \& \ref{fig:modelpvs}).
These internal shocks are so strong that the molecules in
the newly shocked fast wind 
in the inner boundary of the shell 
at the end of the lobe
are also dissociated. Thus, unlike model 3, no HV \H2{} 
emission is seen from the newly shocked fast wind at the far end.

\subsection{Model 5: Pulsed Molecular CFW at 1000 K with $A=0.3$}

This model is the same as model 4, but with a smaller variation amplitude
with A=0.3 (i.e., $\triangle v = 90$ \vkm{}). In this case, the internal
shocks are weaker, so that molecules can survive in the internal shocks
except at their tips (Figs. \ref{fig:models}i, j, E, see the contours of $f$). 
Therefore, HV \H2{}
emission is seen arising from the internal shocks, 
with the intensity decreasing rapidly with the distance due to radial expansion
(Fig. \ref{fig:modelmoms2com}).
It is seen arising from the first three internal shocks, forming
three HV emission clumps (or knots) that
may correspond to the three HV \H2{} emission peaks seen in the \H2{}
observations (Fig. \ref{fig:momobs}).
The HV \H2{} emission seen at $\sim$ 2" away from the source can be compared
to that seen at similar distance in the observations. However, its kinematics, with the velocity
increasing toward the tip and an emission peak at the highest velocity (Fig.
\ref{fig:modelpvs}), is
inconsistent with the observations (Fig. \ref{fig:pvobs}). 
HV CO J=6-5 emission is seen from the wings of the internal shocks and
may show similar PV structure to that seen in the observations. 
However, it is too weak (relative to the LV CO J=6-5 emission) 
to explain the observations.
In addition, the
internal shocks are too hot to have CO J=2-1 emission.

\subsection{Model 6: Pulsed Molecular CFW at 1000 K with $A=0.1$}

This model is the same as model 5, but with a smaller variation amplitude
with A=0.1 (i.e., $\triangle v = 30$ \vkm{}). In this case, the internal
shocks become too weak to dissociate any molecules in the internal
shocks (Figs. \ref{fig:models}k, l).
Molecules also can survive in 
the newly shocked fast wind in the inner boundary of the shell at
the far end of the lobe, producing HV \H2{} emission at the far end as in model
3. The internal shocks
are cold enough to have HV CO emission in J=6-5, but too cold to have HV
\H2{} emission and too hot to have HV CO emission in J=2-1. In addition, due to
radial expansion, the intensity of the HV CO emission in J=6-5 at $\sim$ 2"
is still too faint to be compared with the LV component (Fig.
\ref{fig:modelpvs}), inconsistent with
the observations (Figs. \ref{fig:momobs} and \ref{fig:pvobs}). Moreover,
there will be no atomic and ionic emission from the internal shocks,
inconsistent with the optical observations that show atomic and ionic
emission along the body of the lobe.

\subsection{Summary of our models}

In summary, limb-brightened shell structures are seen in our models in CO
and \H2{} at low velocity arising from the shocked AGB wind in the shell,
and can be identified as the LV components in the observations. 
However, the shell structure in CO J=2-1 is significantly less extended than
that seen in the observations. None of our models can properly reproduce
the observed HV molecular emission near the source along the body of the
lobe. In our models in which the CFWs are atomic, the shell
has high-velocity molecular material only at the far end of the lobe.
Thus, in order to produce HV molecular emission near the source,
the CFW itself has to be molecular.
In our steady CFW models, although
there is some HV CO emission near the source from the CFW itself, it is
far too weak (relative to the LV CO emission) to explain the observations.
This is because the column densities of the HV molecular material decrease
very rapidly due to radial expansion.
In our pulsed CFW models, HV
\H2{} emission can be seen along the outflow axis
arising from the internal shocks and may correspond to that seen in the
observations. However, its kinematics is inconsistent with
the observations. 
HV CO 6-5 emission can be seen arising from the internal shocks and
may show similar PV structure to that seen in the observations. 
However, it is too weak (relative to the LV CO J=6-5 emission) 
to explain the observations.
In addition, HV CO J=2-1 emission from the internal shocks
is far too weak.

\section{Discussion}

\subsection{Could the HV component be from a different lobe?}

In high-resolution optical images, CRL 618 is clearly seen with multiple
lobes on each side and some of these lobes are not as extended as the W1
lobe \cite[see Fig. \ref{fig:momobs} and][]{TG02}. It is thus possible that
the HV component is actually associated with a small lobe that happens to be
aligned with the W1 lobe, and arises from the shell instead of the internal
shocks. This small lobe could be either a highly inclined extended lobe that
appears small in projection or a younger lobe that has a smaller physical
linear extent compared to the W1 lobe.  Our model 3, which has molecular
emission from both the shocked AGB wind and shocked fast wind in the shell,
can be used to investigate this possibility.

At a higher inclination of
60\degree{}, the outflow lobe has a projected length of $\sim$ \arcs{3}.
At this higher inclination, the emission from the shocked AGB wind
has a higher velocity, but still much lower than the observed HV component
(Fig. \ref{fig:modelpvsHS}). The two HV \H2{} emission from the
shocked fast wind at the far end are now projected to
$\sim$ \arcsa{2}{5} at $\sim$ 100 \vkm{} and $\sim$ \arcs{3} at $\sim$
180 \vkm{}, respectively.
The one at \arcsa{2}{5} can be compared with the
observed \H2{} component at the same distance.
However, its velocity increases
with the distance, inconsistent with the observations. In
addition, no HV CO emission is seen from the shocked fast wind. Only faint
HV CO emission is seen near the source from the CFW itself.

At a younger age of 86 years, the outflow lobe also has a projected length
of $\sim$ \arcs{3} at the inclination of 30\degree{}. The emission from the shocked
AGB wind and shocked fast wind has the same velocity structure as at older
age (Fig. \ref{fig:modelpvsHS}), because the shell is momentum-driven and thus roughly self similar. 
The HV \H2{} emission from the shocked fast wind at $\sim$ \arcs{2} may show
a similar PV structure to that seen at $\sim$ \arcsa{2}{5} in the
observations.  However, no HV CO emission is seen from the shocked fast
wind. Again, only faint HV CO emission is seen near the source from the CFW
itself.

As a result, 
adding a small lobe into our models is
still not able to reproduce the observed HV CO and \H2{} components in
CRL 618 properly.
The HV molecular gas in CRL 618 might have a layered structure, with
different temperatures and thus different
emissions in different layers.

\subsection{Could the CFW have a different physical structure?}

The CFW may have a different physical structure, 
as concluded in \citet{LS03} when comparing optical emission along the body
of the lobe in our models
to the observations. Here we discuss the two possible physical structures of the
CFW that have been studied in the literature.

\subsubsection{Episodic Cylindrical jet?}

The CFW could be in a form of a cylindrical jet, with the fast wind material
confined to a small cross section and collimated to the same direction along
the outflow axis. In this case, the CFW has a constant density with the
distance from the source. The observations
in CO J=2-1 also show that the HV component is better reproduced by a
cylinder with the gas flowing axially \citep{Sanchez2004}.
Simulations with episodic
cylindrical jets have been performed by a number of authors 
in order to reproduce the morphology and kinematics of protostellar outflows 
\cite[e.g.,][]{Suttner1997,Lee2001}. 
Such jets can also produce a series of
internal bow shocks with HV CO and \H2{} emissions
along the body of the lobe \citep{Suttner1997}.
Without radial expansion, the emission of
these bow shocks does not decrease as fast as that in our current models.
In addition, unlike those in our current models, the shocks are ballistic
\citep{Lee2001}.
The cylindrical jet models have been applied to PNs
\citep{Cliffe95,Steffen98,LS04,Guerrero2008} and may apply to
PPNs as well.

The jet is likely to be magnetized because the magnetic field can provide
the required collimation. In addition, a magnetized jet has also been found
to better reproduce the jet emission in the young PN Hen 2-90 than an
unmagnetized jet \citep{LS04}.
The jet could be launched by magneto-centrifugal forces
from a magnetized accretion disk and star system \citep{Blackman2001}, like
the protostellar jets \citep{Shu1994,Konigl2000}. 
The central star could have a binary companion and
the accretion disk could form as the material flows
to the companion \citep{Mastrodemos1998}.

\subsubsection{Bullet?}

Recently, a bullet (i.e., a massive clump) model has also been proposed for
PPN shaping because it can also produce a collimated outflow lobe
\citep{Dennis08}. The bullet could be launched by an explosive MHD mechanism
\cite[see, e.g.,][]{Matt2006}. It is not clear, however, how a bullet can
produce HV molecular emission near the source along the body of
the lobe. One way that we can think of is to have a
chain of bullets along the body of the lobe.
Note that, however, a jet model with a periodic variation in
velocity has been found to be better
than a model with a chain of bullets or dense clumps, 
in reproducing the knotty jet emission in the young PN Hen
2-90 \cite[e.g., comparing models 1 and 2 in][]{LS04}.

\citet{Dennis08}, having compared their jet and bullet models,
argued that the bow-shock heads of bullets take on a V-shaped configuration
and are thus more consistent with the observations of CRL 618, whereas
bow-shock heads of jets are more U-shaped. However, the assumed density and thus the
mass-loss rate of the AGB
wind in their models are about a factor of 1000 less than those
assumed in our
models, which are more appropriate for CRL 618 \citep{Knapp1985,Sanchez2004}.
With a high density of the AGB wind as observed, the jet model has been
found to produce V-shaped heads \cite[see, e.g.,][]{Suttner1997,Lee2001}
similar to that seen in CRL 618, because of efficient cooling at high
density. The jet model in \citet{Dennis08} may instead apply to another PPN
OH 231.8+4.2, which shows a wide and more U-shaped outflow lobe
\citep{Bujarrabal2002}.

\subsection{Is the CFW molecular?}

In our models, the CFW is assumed to be molecular in order to have HV
molecular emission near the source along the body of the lobe. It could
be intrinsically molecular. It could also be intrinsically atomic and become
molecular right after launched due to its high mass-loss rate, as proposed 
for protostellar wind \citep{Glassgold1991}.
Alternatively, the HV molecular emission could arise from the entrained AGB
material that was originally close to the source, as suggested by \citet{Cox03},
or in the inner
dense torus-like core around the source, as suggested by \citet{Sanchez2004}.
It is not clear in our simulations, however,
how this entrainment
can happen once the shell is formed and shields the fast wind from
being interacting with the AGB wind around the source. 
It probably can happen if the fast wind is actually launched from around a 
binary companion.
As the AGB wind flows to the companion, an accretion disk can form 
\citep{Mastrodemos1998} and launch the fast wind, and then the fast wind can
entrain the AGB wind. 

\section{Conclusions}

We have presented simulations of CFW models including molecular cooling and
time dependent chemistry of hydrogen. 
We have also derived from our simulations
intensity maps
and position velocity diagrams for
CO J=2-1, J=6-5, and \H2{} 1-0 S(1) emission
and compared them to recent observations of CRL 618.
In our models, limb-brightened shell structures are seen in CO and \H2{} at
low velocity arising from the shocked AGB wind in the shell, and can be
identified as the low-velocity (LV) components in the observations.
However,
the shell structure in CO J=2-1 is significantly less extended than that
seen in the observations. 
None of our models can properly
reproduce the observed high-velocity (HV) molecular
emission near the source along the body of the lobe.  In order to
reproduce the HV molecular emission, the CFW is required to have
a different structure. One possible CFW structure is the cylindrical jet,
with the fast wind material confined to a small cross section and collimated
to the same direction along the outflow axis.

\acknowledgements
C.-F. Lee and M.-C. Hsu are financially supported by the NSC grant
NSC96-2112-M-001-014-MY3. RS thanks NASA for funding this work via an LTSA
award (NMO710840-102898); RS also received partial support for this work
from an HST GO award (no. GO-10317.01) from the Space Telescope Science
Institute (operated by the Association of Universities for Research in
Astronomy, Inc., under NASA contract NAS5-26555). Some of the research
described in this paper was carried out by RS at the Jet Propulsion
Laboratory, California Institute of Technology, under a contract with the
National Aeronautics and Space Administration.


\hspace{-2in}
\begin{deluxetable}{lccc}
\tabletypesize{\normalsize}
\tablecaption{Physical parameters of the CFWs \label{tab:models}}
\tablewidth{0pt}
\tablehead{
\colhead{Model} & \colhead{Temperature (K)} & \colhead{Composition} &
\colhead{Status}
}
\startdata
1 & 10$^4$ & Atomic    &Steady              \\
2 & 10$^4$ & Atomic    &Pulsed with $A=0.5$ \\
3 & 10$^3$ & Molecular &Steady              \\
4 & 10$^3$ & Molecular &Pulsed with $A=0.5$ \\
5 & 10$^3$ & Molecular &Pulsed with $A=0.3$ \\
6 & 10$^3$ & Molecular &Pulsed with $A=0.1$ \\
\enddata
\end{deluxetable}

\clearpage

\setcounter{figure}{2} 
\begin{figure}
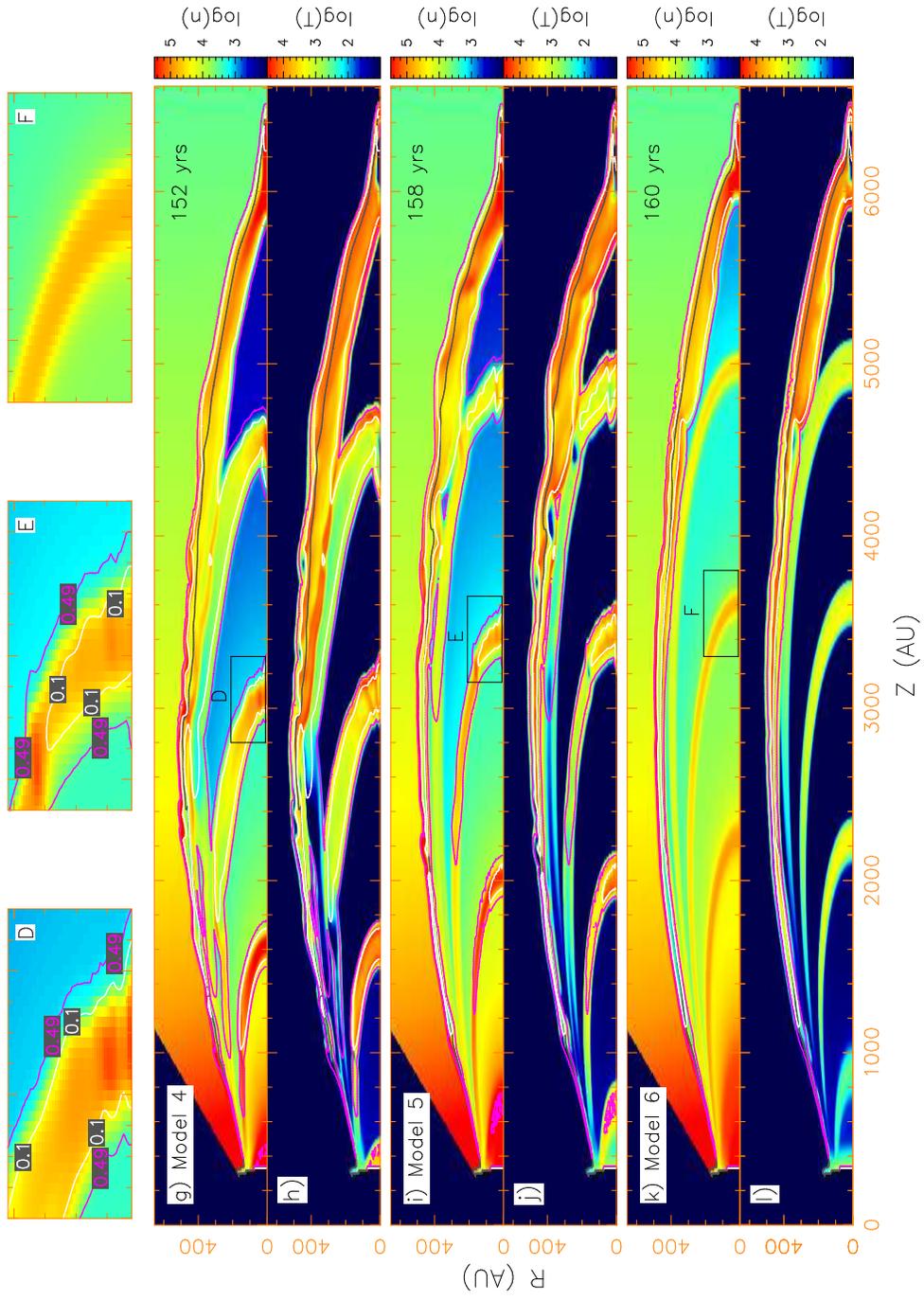

\centering 
\putfig{0.75}{0}{f3b.ps}
\figcaption[] 
{\textit{Continued}
}
\end{figure}
\clearpage

\begin{figure}
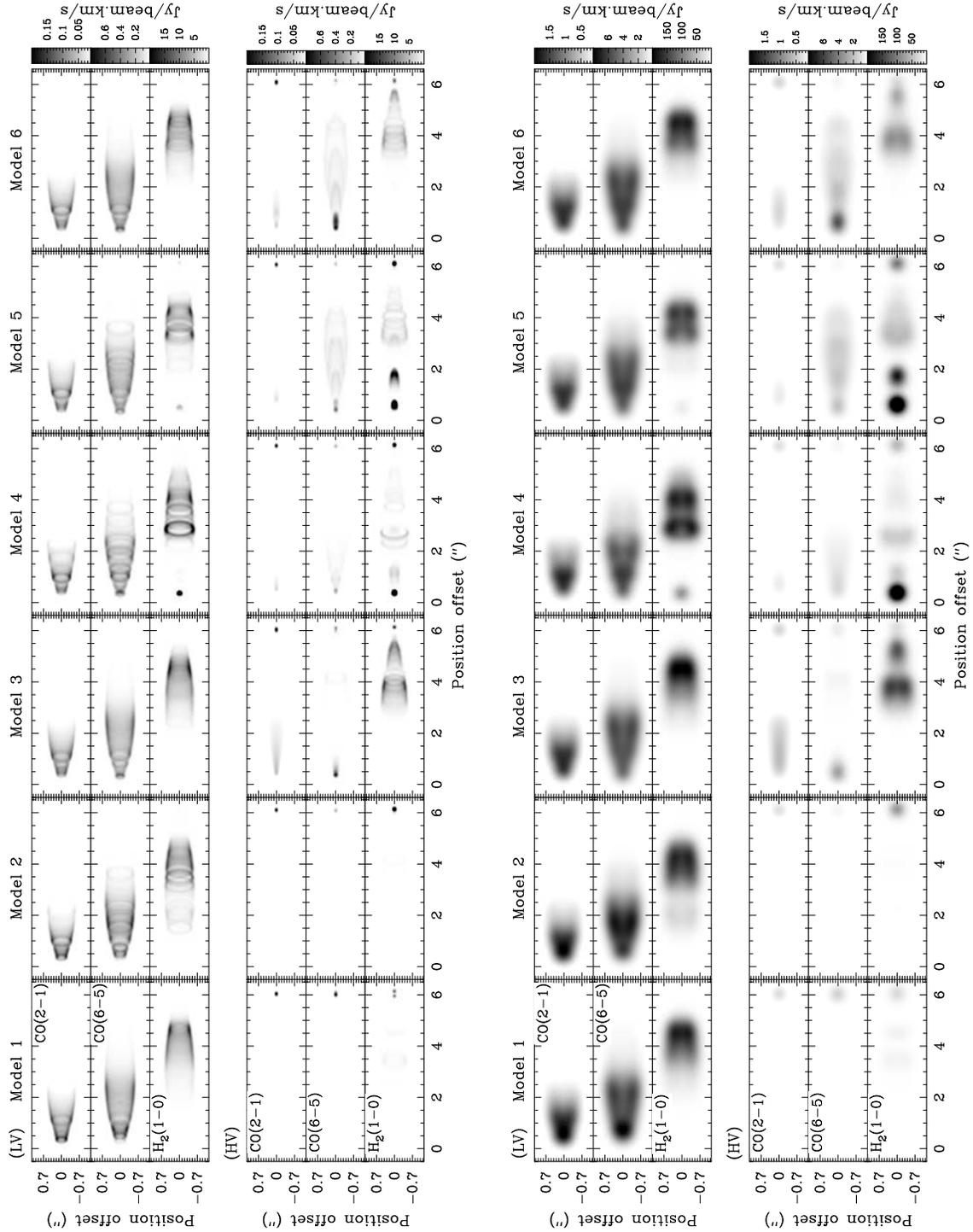

\centering
\putfig{0.75}{0}{f4a.ps}
\mbox{}\hspace{4mm}
\putfig{0.75}{0}{f4b.ps}
\figcaption[]{
Integrated CO J=2-1 (top row), CO J=6-5 (middle row),
and \H2 {} (bottom row) intensity maps derived
from our models 1 to 6 from left column to right column.
The first set and second set of panels show the LV and HV components at
\arcsa{0}{1} resolution, respectively.
The third set and fourth set of panels show the LV and HV components at
\arcsa{0}{5} resolution, respectively, for comparing to the observations.
The HV components are derived by integrating the
emission with velocity higher than 50 \vkm{} from the systemic and the LV
components by integrating the emission with velocity within 50 \vkm{} from
the systemic. 
\label{fig:modelmoms2com} 
}
\end{figure}

\begin{figure}
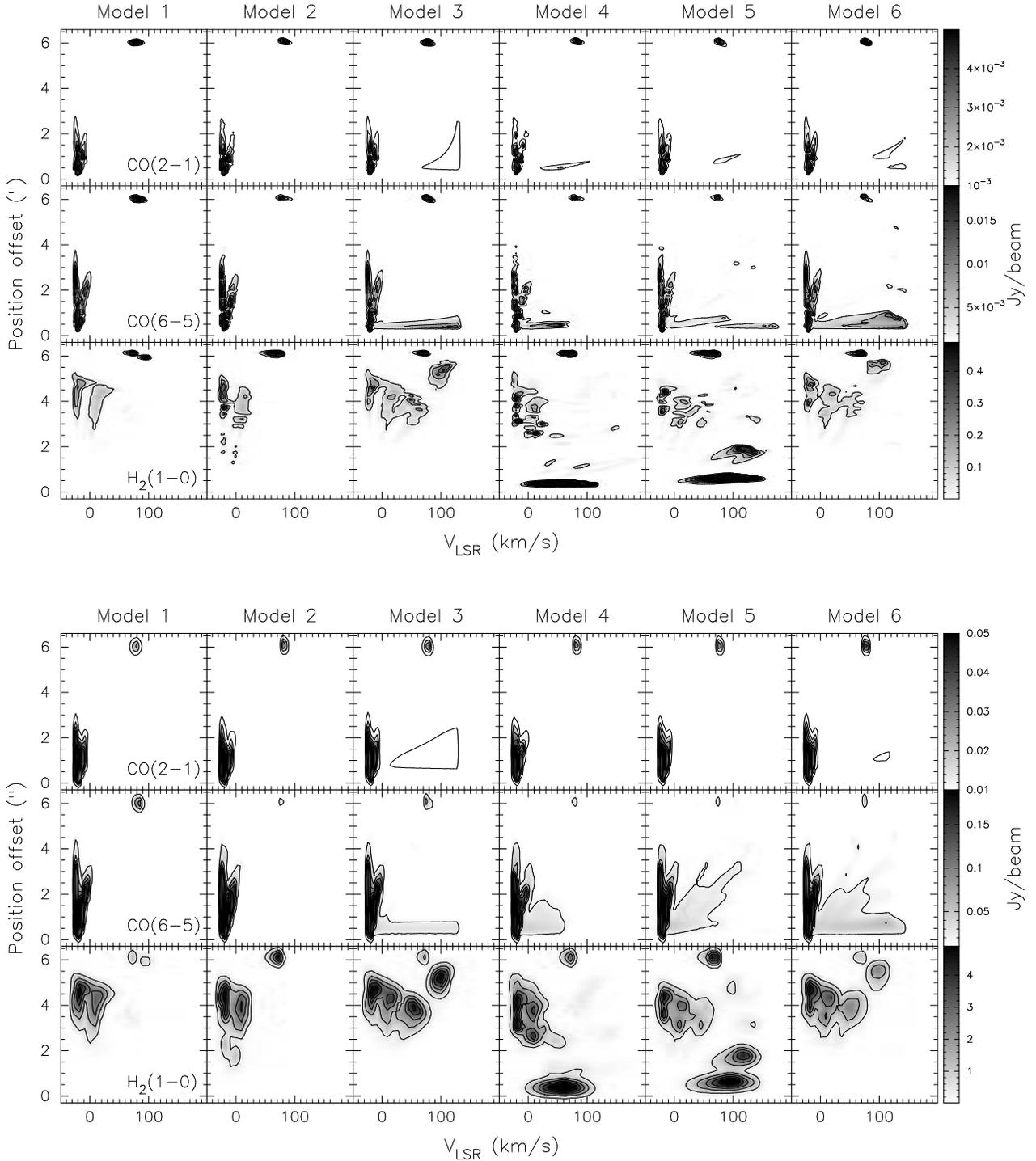

\centering
\putfig{0.65}{270}{f5a.ps}
\mbox{}\vspace{3mm}
\putfig{0.65}{270}{f5b.ps}
\figcaption[]{
PV diagrams of CO J=2-1 (top row), CO J=6-5 (middle row), and \H2 {} (bottom
row) emission cut along the outflow axis, derived from our models 1 to 6.
Top set is derived with a spatial
resolution of \arcsa{0}{1} and a velocity resolution of 4 \vkm{}. Bottom
set is derived with a spatial resolution of \arcsa{0}{5} and a
velocity resolution of 5.2 \vkm{} for CO and 9 \vkm{} for \H2{} emission, for
comparing with the observations.
The contours go from 10\% to 90\% with a step of 20\% of the peak value.
\label{fig:modelpvs} 
}
\end{figure}

\begin{figure}
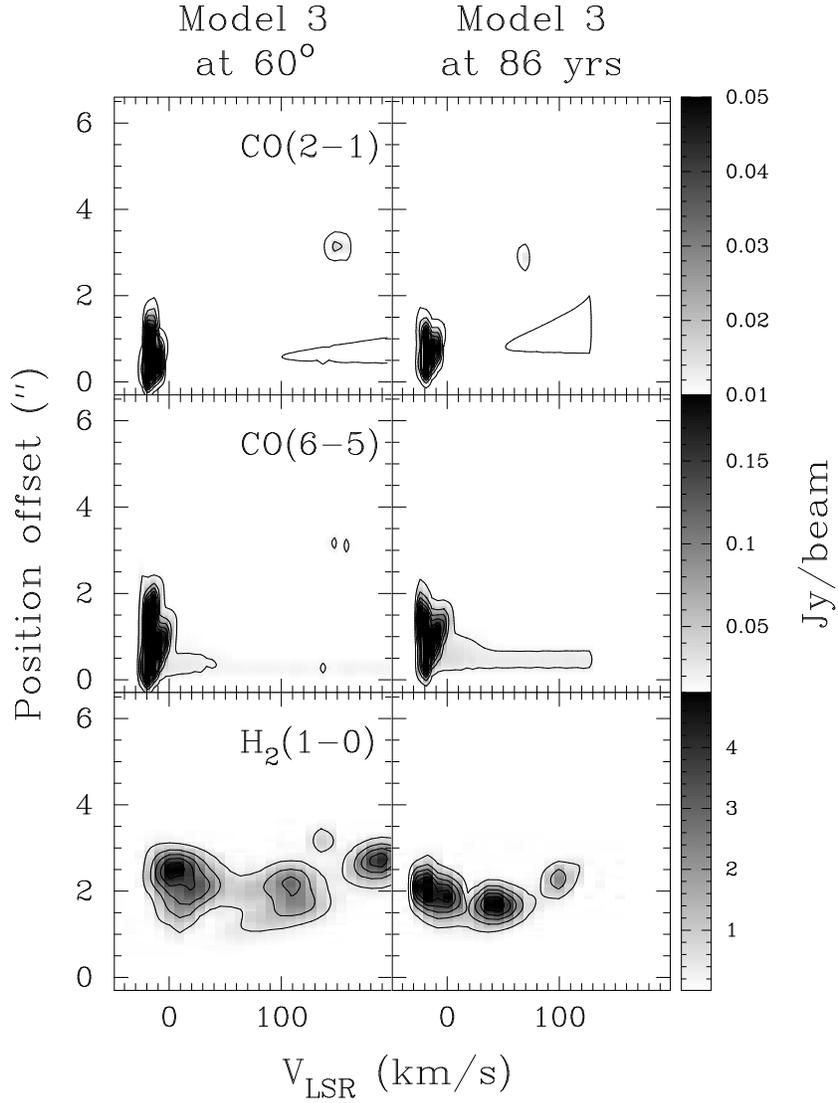

\centering
\putfig{1}{270}{f6.ps}
\figcaption[]{
PV diagrams of CO J=2-1 (top row), CO J=6-5 (middle row), and \H2 {} (bottom
row) emission cut along the outflow axis, derived from our model 3 at a higher
inclination of 60\degree{} (left column) and a younger age of 86 yrs
(right column). They are derived with a spatial resolution of \arcsa{0}{5} and a
velocity resolution of 5.2 \vkm{} for CO and 9 \vkm{} for \H2{} emission, for
comparing with the observations.
The contours go from 10\% to 90\% with a step of 20\% of the peak value.
\label{fig:modelpvsHS} 
}
\end{figure}

\end{document}